\documentclass[11pt]{article}

\usepackage{microtype}
\usepackage{graphicx}
\usepackage{subcaption}
\usepackage{booktabs}
\usepackage{microsoft-tech-report}
\usepackage{hyperref}

\usepackage{xcolor}
\usepackage{colortbl}

\definecolor{gain}{HTML}{2E7D32} 
\definecolor{loss}{HTML}{C62828} 
\definecolor{liteblue}{HTML}{E3F2FD}

\newcommand{\gain}[1]{{\color{gain}\scriptsize +#1}}

\definecolor{vscodedark}{HTML}{1E1E1E}   
\definecolor{vscodeframe}{HTML}{2D2D2D} 
\definecolor{vscodetext}{HTML}{D4D4D4}  

\usepackage{booktabs}   
\usepackage{amsfonts}  
\usepackage{tabularx}   
\usepackage{enumitem}  
\usepackage{tcolorbox}
\tcbuselibrary{skins}

\newtcolorbox{insightbox}{
    enhanced,
    colback=gray!5,                
    colframe=teal!60!black,        
    arc=6pt,                       
    boxrule=0.8pt,                 
    top=8pt, bottom=8pt,           
    left=12pt, right=12pt,         
    shadow={1.5pt}{-1.5pt}{0pt}{black!20!white}, 
    fontupper=\normalsize 
}

\usepackage{amsmath}
\usepackage{amssymb}
\usepackage{mathtools}
\usepackage{amsthm}
\usepackage{bm}
\usepackage{multirow}
\usepackage{enumitem}
\usepackage[capitalize,noabbrev]{cleveref}
\usepackage{xspace}
\usepackage{titletoc}
\usepackage{placeins}  
\usepackage{tcolorbox} 
\usepackage[ruled,linesnumbered,noend,noline]{algorithm2e} 
\DontPrintSemicolon
\usepackage{listings}  
\usepackage{hyperref}
\usepackage{marvosym}
\tcbuselibrary{theorems,skins,breakable,listings}

\definecolor{promptbg}{HTML}{F7F7F8}
\definecolor{promptborder}{HTML}{5B5B5B}
\newtcolorbox{promptbox}[1][]{
  colback=promptbg, colframe=promptborder, fonttitle=\bfseries\sffamily,
  breakable, enhanced, arc=3pt, boxrule=0.8pt,
  left=10pt, right=10pt, top=8pt, bottom=8pt,
  title={#1}
}

\lstdefinestyle{json}{
  basicstyle=\small\ttfamily,
  breaklines=true,
  breakatwhitespace=false,
  showstringspaces=false,
  columns=fullflexible,
  literate={"}{\textquotedbl}1,
}

\newif\ifappendixtoc
\appendixtoctrue   

\definecolor{color_blue}{HTML}{E7EFFA}
\definecolor{color_green}{HTML}{E6F8E0}
\definecolor{color_gray}{HTML}{ECECEC}
\definecolor{pearDark}{HTML}{2980B9}
\definecolor{theoremblue}{HTML}{EBF5FB}
\definecolor{theoremborder}{HTML}{2980B9}
\definecolor{propgreen}{HTML}{EAFAF1}
\definecolor{propborder}{HTML}{27AE60}
\definecolor{defyellow}{HTML}{FEF9E7}
\definecolor{defborder}{HTML}{F39C12}
\definecolor{remarkgray}{HTML}{F2F3F4}
\definecolor{remarkborder}{HTML}{7F8C8D}

\hypersetup{
  colorlinks=true,
  linkcolor=msftblue,
  citecolor=msftblue,
  urlcolor=msftblue,
  pdftitle={SkillOpt-Lite and HarnessOpt}
}

\newtcbtheorem[auto counter]{theorem}{Theorem}{
  colback=theoremblue, colframe=theoremborder, fonttitle=\bfseries,
  breakable, enhanced, sharp corners, boxrule=0.6pt, left=6pt, right=6pt, top=6pt, bottom=6pt,
  crefname={Theorem}{Theorems}
}{thm}

\newtcbtheorem[use counter from=theorem]{proposition}{Proposition}{
  colback=propgreen, colframe=propborder, fonttitle=\bfseries,
  breakable, enhanced, sharp corners, boxrule=0.6pt, left=6pt, right=6pt, top=6pt, bottom=6pt,
  crefname={Proposition}{Propositions}
}{prop}

\newtcbtheorem[use counter from=theorem]{lemma}{Lemma}{
  colback=theoremblue, colframe=theoremborder, fonttitle=\bfseries,
  breakable, enhanced, sharp corners, boxrule=0.6pt, left=6pt, right=6pt, top=6pt, bottom=6pt,
  crefname={Lemma}{Lemmas}
}{lem}

\newtcbtheorem[use counter from=theorem]{corollary}{Corollary}{
  colback=propgreen, colframe=propborder, fonttitle=\bfseries,
  breakable, enhanced, sharp corners, boxrule=0.6pt, left=6pt, right=6pt, top=6pt, bottom=6pt,
  crefname={Corollary}{Corollaries}
}{cor}

\newtcbtheorem[use counter from=theorem]{definition}{Definition}{
  colback=defyellow, colframe=defborder, fonttitle=\bfseries,
  breakable, enhanced, sharp corners, boxrule=0.6pt, left=6pt, right=6pt, top=6pt, bottom=6pt,
  crefname={Definition}{Definitions}
}{def}

\newtcbtheorem[use counter from=theorem]{assumption}{Assumption}{
  colback=defyellow, colframe=defborder, fonttitle=\bfseries,
  breakable, enhanced, sharp corners, boxrule=0.6pt, left=6pt, right=6pt, top=6pt, bottom=6pt,
  crefname={Assumption}{Assumptions}
}{asm}

\newtcbtheorem[use counter from=theorem]{remark}{Remark}{
  colback=remarkgray, colframe=remarkborder, fonttitle=\bfseries,
  breakable, enhanced, sharp corners, boxrule=0.6pt, left=6pt, right=6pt, top=6pt, bottom=6pt,
  crefname={Remark}{Remarks}
}{rem}

\crefname{tcb@cnt@theorem}{Theorem}{Theorems}
\crefname{tcb@cnt@proposition}{Proposition}{Propositions}
\crefname{tcb@cnt@lemma}{Lemma}{Lemmas}
\crefname{tcb@cnt@corollary}{Corollary}{Corollaries}
\crefname{tcb@cnt@definition}{Definition}{Definitions}
\crefname{tcb@cnt@assumption}{Assumption}{Assumptions}
\crefname{tcb@cnt@remark}{Remark}{Remarks}
\crefname{algocf}{Algorithm}{Algorithms}

\techreportshorttitle{Mage-VL}

\begin{document}
\thispagestyle{empty}

\noindent
\begin{minipage}[c]{0.5\linewidth}
\raggedright
\raisebox{-0.5\height}{\msftbrandmark}
\end{minipage}
\begin{minipage}[c]{0.49\linewidth}
\raggedleft
{\msftdatefont\small\color{msftgray}July, 2026}    
\end{minipage}\par
\vspace{0.35em}
\noindent{\color{msftline}\rule{\linewidth}{0.8pt}\par}

\vspace{1.0em}
\begin{center}
{{\msfttitlefont\fontsize{21}{25}\selectfont\color{msftdark}
SkillOpt-Lite: Better and Faster Agent Self-evolution via One Line of Vibe\par}}
\vspace{1.25em}

{\normalsize\msftsans\bfseries\color{msftdark}
Yifei Shen$^{1}$ \quad Bo Li$^{1,2}$\quad Xinjie Zhang$^{3}$ \par}
\vspace{0.8em}

{\normalsize\rmfamily\color{msftdark}
$^{1}$LMMs-Lab \quad $^{2}$NTU MMLab \quad $^{3}$Microsoft\par
}
\end{center}

\vspace{0.45em}
\begin{msfttitlebox}
\setlength{\parindent}{0cm}
\setlength{\parskip}{0cm}

\begin{abstract}
While skill optimization for autonomous agents has gained traction, existing methods rely on complex pipelines. This leaves a fundamental question unaddressed: What constitutes a minimal viable pipeline for skill optimization, where every component is justified by theory or empirical necessity? We formalize skill optimization via Zeroth-Order (ZO) optimization, mapping classical counterparts (central difference, trust regions) to recent literature. Noting that unlike blind numerical perturbations in classical ZO, skill trajectories serve as interpretable debugging feedback. Grounded in Claude Code philosophy and PAC learning, we establish three principles for convergence and generalization: file-system-based trajectory exploration, consensus attribute mining, and independent validation gating. Eliminating redundancies, we propose SkillOpt-Lite. It accelerates convergence and outperforms full SkillOpt: improving LiveMath by +8.8 points on GPT-5.5 and +25.4 points on GPT-5.4-nano, allowing the nano model to surpass standard GPT-5.4 optimized by SkillOpt. Finally, we integrate our framework into production coding agents like VSCode Copilot, enabling developers to evolve agent skills via one line of vibe. Because our framework treats all agent components simply as standard editable code, this minimal pipeline naturally generalizes to full harness optimization (HarnessOpt). On SpreadsheetBench, HarnessOpt enables GPT-5.4-nano to achieve 0.7758 accuracy, outperforming the larger GPT-5.5 running standard pipelines (0.7620). Code is available at https://github.com/EvolvingLMMs-Lab/SkillOpt-Lite.
\end{abstract}

\vspace{0.14cm}
{\setlength{\parskip}{0.06cm}\small
}

\end{msfttitlebox}
\suppressfloats[t]  


\begin{figure*}[htbp]
    \centering
    \begin{subfigure}[b]{0.46\textwidth}
        \centering
        \includegraphics[width=\textwidth]{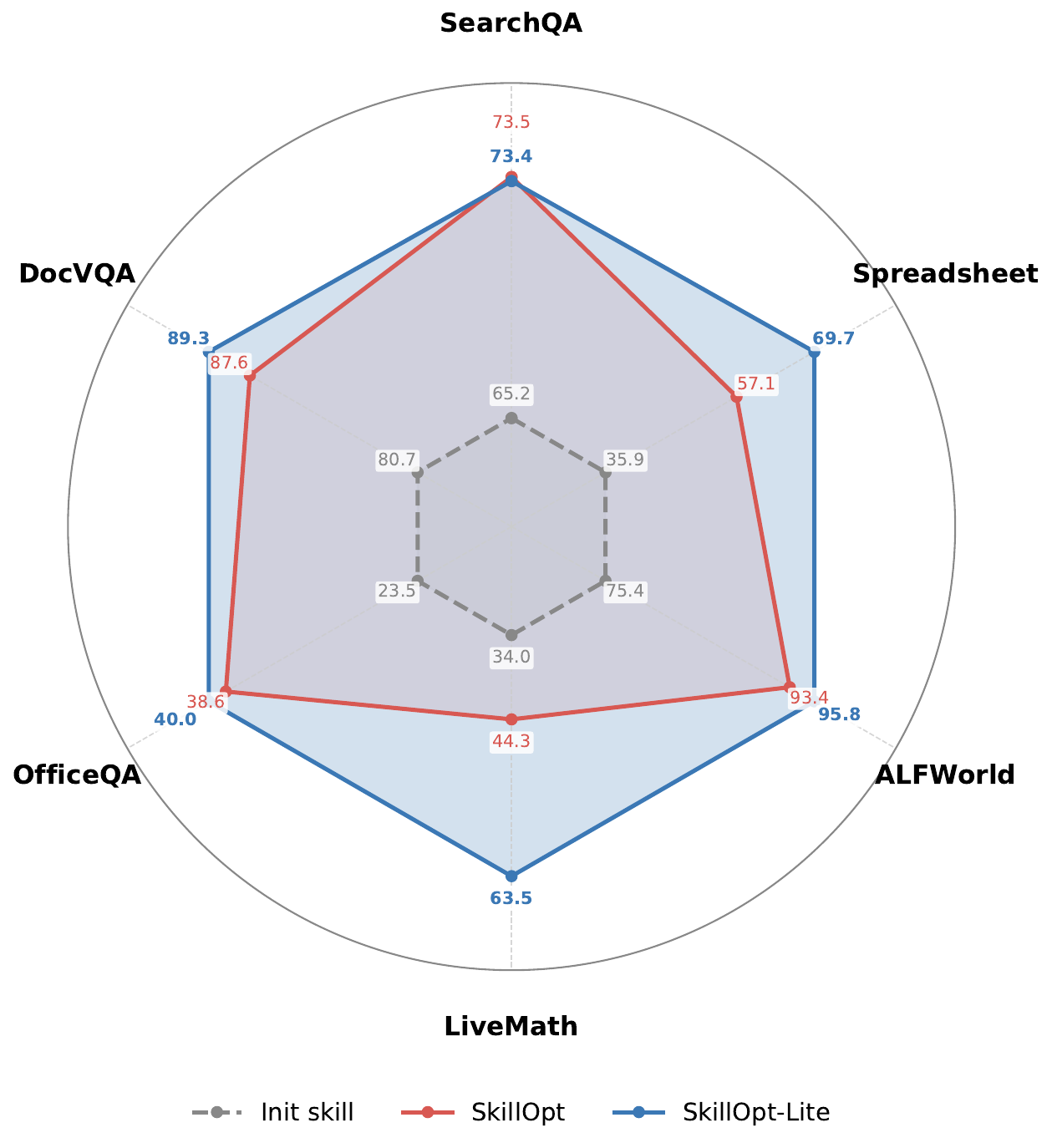}
        \caption{Macro-level average performance profiling.}
    \end{subfigure}
    \hfill
    \begin{subfigure}[b]{0.51\textwidth}
        \centering
        \includegraphics[width=\textwidth]{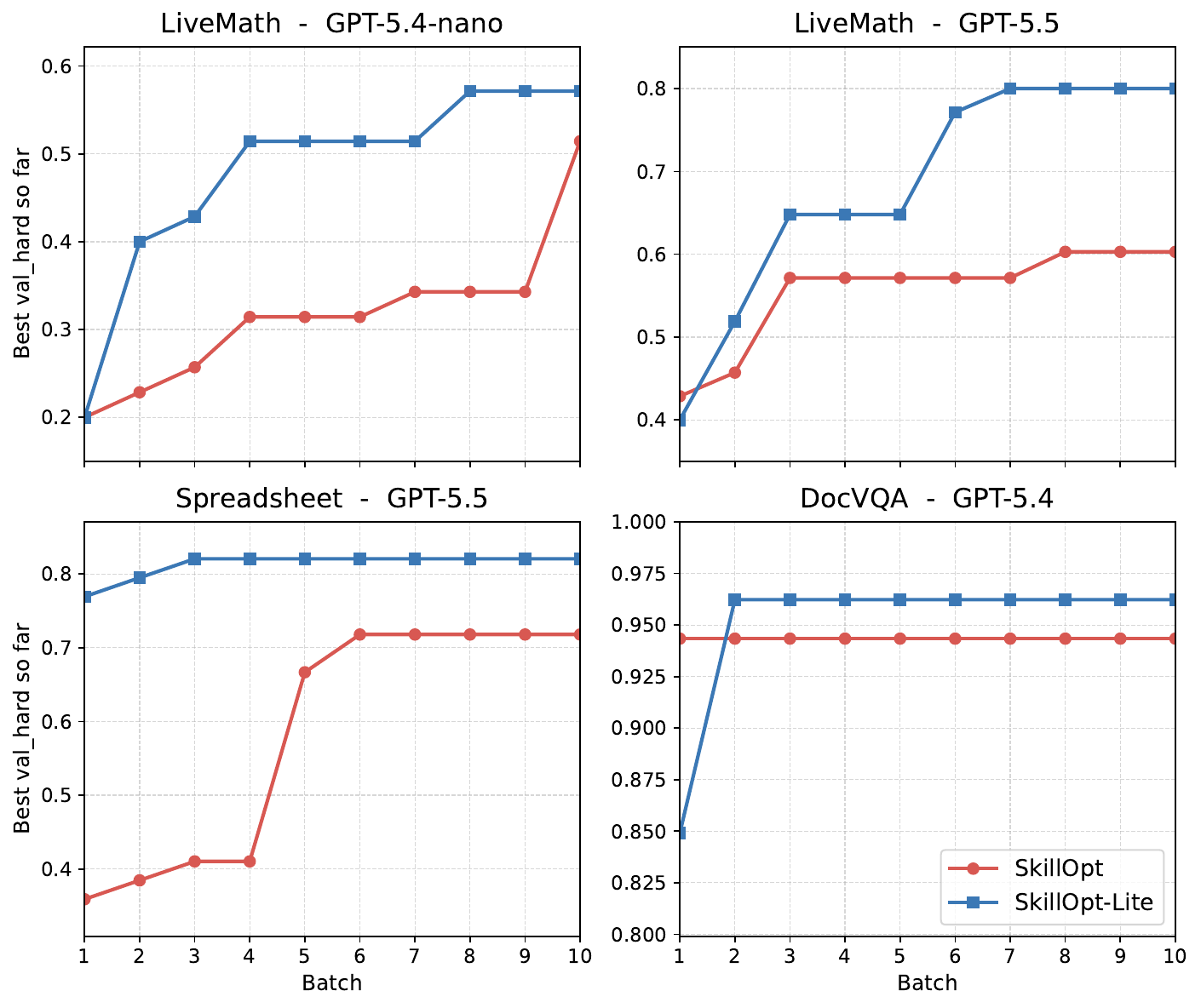}
        \caption{Micro-level validation convergence trajectories.}
    \end{subfigure}
    \caption{The overall performance profile and optimization velocity of SkillOpt-Lite. (a) The radar plot illustrates the macro-level capability comparison averaged across all evaluated model scales, showing comprehensive performance gains over both the initial skills and the original SkillOpt. (b) The micro-level convergence curves on representative tracking segments demonstrate that our minimal viable pipeline delivers significantly accelerated policy refinement and higher validation ceilings (\texttt{Best val\_hard so far}) within a tight 10-batch horizon.}
    \label{fig:teaser_main}
\end{figure*}

\section{Introduction}
\label{sec:introduction}

AI agents have transitioned from research prototypes into standard production utilities~\cite{anthropic2025claudecode,openclaw2025,nous2026hermes}. Practical deployment reveals that an agent's real-world capability is determined not solely by its base Large Language Model (LLM), but by the interaction among the base model, its operational scaffolding (harness), and its domain heuristics (skills). Because foundation models remain frozen during inference, agent engineering frequently reduces to refining these skill documents—a process where subtle textual variations routinely drive non-linear disparities in downstream task performance~\cite{yang2026skillopt,yang2024swe,lee2026meta,lin2026agentic,ni2026trace2skill}.

To manage this non-linear sensitivity, automated skill optimization has evolved from open-loop self-reflections~\cite{shinn2023reflexion,wang2023voyager} into complex algorithmic pipelines~\cite{ni2026trace2skill,yu2026skilladaptor,liu2026skillforge,chen2026skillcat}. Among existing literature, \textit{SkillOpt}~\cite{yang2026skillopt} formalizes text revision through continuous optimization analogies, implementing mini-batch tree merging, textual learning-rate schedules, and rejected-edit buffers. Yet, despite their empirical gains, existing frameworks exhibit a growing tendency toward architectural complexity. They incorporate intricate mechanics from classical optimization without addressing a fundamental question: 
\begin{quote}
    \centering
    What constitutes a minimal viable pipeline for skill optimization, where every component is justified by theoretical or empirical necessity?
\end{quote}

In this report, we address this question by bridging zeroth-order optimization, statistical learning theory, and systems engineering. We first map existing reflection paradigms to Zeroth-Order (ZO) optimization. As detailed in Table~\ref{tab:zo_agent_mapping}, single-trace critiques~\cite{shinn2023reflexion,wang2023voyager} correspond to 1-point gradient estimators $\hat{\nabla}f(s)$; contrastive diagnoses~\cite{chen2026skillcat} reflect central difference approximations; fault-isolated edits~\cite{yu2026skilladaptor} mirror coordinate descent along basis vectors $e_i$; while edit budgets~\cite{yang2026skillopt,liu2026skillforge} and rejected buffers instantiate trust region radii $\mathcal{B}$ and control variates $c$. However, we identify a key conceptual divergence: classical ZO relies on blind numerical perturbations over unobservable states, whereas agentic rollouts yield explicit, semantic-rich execution traces. Skill optimization functions fundamentally as language-mediated program compilation where trajectories serve as interpretable debugging feedback.

To formalize this compiler paradigm, we establish three foundational design principles. The first two emerge from PAC-learning theory:
\begin{enumerate}[leftmargin=*, label=(\arabic*)]
    \item \textbf{Consensus Mining across Trajectories:} Overfitting to single-trial anomalies inflates the stability coefficient $\beta_{\exp}$, causing generalization collapse. Instead of executing complex tree-reductions, the optimizer must act as a compression operator capturing cross-task invariants.
    \item \textbf{Independent Validation Gating:} PAC bounds demonstrate that a held-out validation set removes $\beta_{\exp}$ from the generalization error. To prevent evaluation bias, gating must rely on independent samples rather than sub-sampled training failures.
\end{enumerate}
Guided by the Unix and Claude Code~\cite{claudecode} philosophy of ``everything is a file'', we design a pilot experiment that isolates each execution trajectory into a standalone flat file and deploys an autonomous coding agent to refine them using primitive file-system tools; empirically, we discover that this single-batch file modification may outperform the full SkillOpt pipeline optimized over four full epochs across multiple benchmarks, leading us to derive our third empirical principle:
\begin{enumerate}[leftmargin=*, resume, label=(\arabic*)]
    \item \textbf{A Bitter Lesson of Skill Optimization:} As base LLMs scale, bespoke topologies designed to damp update velocity become counterproductive. Granting agents primitive shell utilities to directly inspect raw log files consistently outperforms heavily engineered baselines.
\end{enumerate}

Synthesizing these principles, we propose \textbf{SkillOpt-Lite}, a minimal viable pipeline that removes mini-batch reflection pooling, slow update damping, and rejection buffers. Instead, \textit{SkillOpt-Lite} operates directly on the disk via an autonomous debugging loop consisting solely of trajectory exploration, consensus mining, and validation gating. As shown in Figure~\ref{fig:teaser_main}, this streamlined design delivers consistent improvements over the fully configured \textit{SkillOpt} baseline across all six benchmarks. On logic-intensive tasks such as Spreadsheet, it yields an average score increase of \textbf{+12.6 points} (69.7 vs. 57.1). Furthermore, as tracked in Figure~\ref{fig:teaser_main}b, removing reflection pooling allows lightweight models to circumvent early exploration plateaus; for instance, on ALFWorld, \textit{SkillOpt-Lite} lifts GPT-5.4-nano to \textbf{81.3\%} (+9.5 over SkillOpt), while demonstrating rapid convergence within the first step on flagship bases (e.g., Spreadsheet-GPT-5.5).

Finally, we encapsulate our pipeline into an open-source VS Code Copilot extension, allowing developers to trigger skill optimization via a single slash command[cite: 1]. Crucially, because our framework treats all agent components simply as standard editable code, lifting file-path restrictions within the IDE environment allows this minimal loop to naturally generalize to full harness optimization—a framework we designate as \textbf{HarnessOpt}. On \texttt{SpreadsheetBench}, \textbf{HarnessOpt} consistently resolves model-specific bottlenecks like reasoning loops. Notably, this framework allows the lightweight \texttt{GPT5.4-nano} to achieve an accuracy of $0.7758$, outperforming the much larger \texttt{GPT5.5} model running under a standard harness and full SkillOpt pipeline ($0.7620$). In summary, our main contributions are threefold:
\begin{itemize}[leftmargin=*]
    \item \textbf{Theoretical Formalization:} We map agentic skill training to zeroth-order optimization, identifying its unique program-compiler nature, and derive core principles governing stability and validation.
    \item \textbf{Minimal Viable Pipeline:} Guided by the file-centric philosophy, we propose \textbf{SkillOpt-Lite}, demonstrating empirically that replacing complex optimization modules with primitive file exploration yields faster convergence and superior performance across multiple baselines.
    \item \textbf{Harness Extension \& IDE Encapsulation:} We show that this pipeline directly extends to automated harness optimization (\textbf{HarnessOpt}), enabling lightweight models to surpass frontier-class models through environment co-design. We encapsulate the entire workflow into a VSCode extension for one-line agent evolution.
\end{itemize}
\section{Analysis of Skill Optimization}
\label{sec:analysis}

\subsection{Skill Training and Zeroth-Order Optimization}
\subsubsection{Algorithmic Connections}
We formalize skill optimization as the maximization of an expected reward function $f(s) = \mathbb{E}_{z \sim \mathcal{D}}[R(H(M, z, s))]$, where $s \in \mathcal{S}_{\text{text}}$ represents a text-based skill artifact, $M$ is a frozen backbone LLM, $z$ denotes a task instance drawn from distribution $\mathcal{D}$, and $H$ signifies the execution harness. Since the gradient $\nabla_s f$ is analytically intractable due to the discrete nature of $\mathcal{S}_{\text{text}}$ and the non-differentiable composition of $H$ and $M$, the agent-environment interaction is modeled as a Zeroth-Order (ZO) Oracle. This formulation reveals that existing heuristics utilize components analogous to the classical ZO toolbox \cite{liu2020primer}, despite the semantic domain's lack of a continuous, smooth manifold. 

Formally, we establish a structural analogy between continuous ZO operations and discrete text space edits: a text artifact $s$ is evaluated at perturbed positions $s + \mu u$ (or along a standard coordinate basis $e_i$), where $\mu > 0$ denotes a step size analogue, and $u$ is a random search direction. Under this unified framework, environmental feedback (e.g., execution traces, error logs, and validation scores) acts as the scalar oracle evaluation $f(s)$. LLM-driven self-reflection, patch generation, and skill editing serve as discrete-space structural analogues to continuous ZO operators, such as stochastic gradient estimators $\hat{\nabla}f(s)$, coordinate descent, trust-region constraints $\mathcal{B}$, and variance-reducing control variates $c$. Consequently, prominent agentic workflows—ranging from single-trace heuristics \cite{shinn2023reflexion, wang2023voyager} to structured multi-point variance-reduction schemes \cite{yang2026skillopt, ni2026trace2skill, liu2026skillforge}—can be systematically analyzed through the lens of algorithmic zeroth-order optimization, as mapped in Table \ref{tab:zo_agent_mapping}.

\begin{table}[!h]
\centering
\footnotesize 
\renewcommand{\arraystretch}{0.85} 
\caption{Mapping Between Zeroth-Order Optimization Operators and LLM/Agent Reflection Paradigms}
\label{tab:zo_agent_mapping}
\begin{tabularx}{\textwidth}{
    >{\raggedright\arraybackslash\bfseries}p{2.6cm} 
    >{\centering\arraybackslash}p{3.5cm} 
    >{\raggedright\arraybackslash}p{3.8cm} 
    >{\raggedright\arraybackslash}X
}
\toprule
\textbf{ZO Concept} & \textbf{Formula / Operator} & \textbf{Agent Domain Metric} & \textbf{Literature Realization} \\
\midrule
Zeroth-Order Oracle & $f(s + \mu u)$ & Sandbox Environment Feedback & Scalar metric of training performance. \\
\midrule
1-Point Estimate & $\hat{\nabla}f(s) \propto f(s+\mu u)u$ & Single-Trajectory Reflection & 
\begin{itemize}[leftmargin=*, nosep, topsep=0pt, partopsep=0pt,before=\vspace{-0.75\baselineskip}]
    \item \textbf{Reflexion} \cite{shinn2023reflexion}: Direct reflection derived from a single trace.
    \item \textbf{Voyager} \cite{wang2023voyager}: Single-error signals triggering local program overwrites.
\end{itemize} \\
\midrule
Multi-Point / Mini-batch & $\frac{1}{b} \sum_{i=1}^{b} [f(s+\mu u_i) - f(s)] u_i$ & Batch Rollout Consensus Extraction & 
\begin{itemize}[leftmargin=*, nosep, topsep=0pt, partopsep=0pt, before=\vspace{-0.75\baselineskip}]
    \item \textbf{Trace2Skill} \cite{ni2026trace2skill}: ZO-SGD using Map-Reduce for patch merging.
    \item \textbf{SkillOpt} \cite{yang2026skillopt}: Iterative reflection mini-batching with size $B_m=8$.
    \item \textbf{SkillForge} \cite{liu2026skillforge}: Phase 2 batch ticket aggregation for trajectory denoising.
\end{itemize} \\
\midrule
Central Difference & $\frac{f(s+\mu u) - f(s-\mu u)}{2\mu}$ & Success-Failure Contrastive Analysis & 
\begin{itemize}[leftmargin=*, nosep, topsep=0pt, partopsep=0pt, before=\vspace{-0.75\baselineskip}]
    \item \textbf{SkillCat} \cite{chen2026skillcat}: Custom CCE operator for trace contrastive analysis at action divergence point $w_i$.
\end{itemize} \\
\midrule
ZO Coordinate Descent & $\frac{f(s+\mu e_i) - f(s)}{\mu} e_i$ & Fault-Isolated Atomic Modification & 
\begin{itemize}[leftmargin=*, nosep, topsep=0pt, partopsep=0pt, before=\vspace{-0.75\baselineskip}]
    \item \textbf{SkillAdapter} \cite{yu2026skilladaptor}: Coordinate descent fixing faulty step $t^*$ as axis and candidate skill $s_j$ as basis vector.
\end{itemize} \\
\midrule
Trust Region Radius & $\mathcal{B}(s_k, \Delta_k)$ & Structural Edit Constraints & 
\begin{itemize}[leftmargin=*, nosep, topsep=0pt, partopsep=0pt, before=\vspace{-0.75\baselineskip}]
    \item \textbf{SkillOpt} \cite{yang2026skillopt}: Edit Budget decay ($L_t = 4 \to 2$).
    \item \textbf{SkillForge} \cite{liu2026skillforge}: Enforcement of the Minimal Modification principle.
    \item \textbf{SoftSkill} \cite{tao2026softskill}: Bounding soft-prefix length at $m=32$ tokens.
\end{itemize} \\
\midrule
Control Variate & $\hat{g}_t - c_t + \mathbb{E}[c]$ & Historical Memory Rejection & 
\begin{itemize}[leftmargin=*, nosep, topsep=0pt, partopsep=0pt, before=\vspace{-0.75\baselineskip}]
    \item \textbf{SkillOpt} \cite{yang2026skillopt}: Passing rejected edits into a buffer as a negative control variate.
\end{itemize} \\
\bottomrule
\end{tabularx}
\end{table}

\subsubsection{Conceptual Divergence}
\label{sec:diff}
Although mapping agentic skill training to ZO optimization introduces established variance-reduction methods, a key conceptual divergence exists between the two paradigms. In classical ZO optimization, the decision variable $s$ is defined within a numerical space. The oracle functions as a strict black box that returns only a scalar reward $f(s)$, while the detailed runtime trajectory or intermediate state transitions remain latent and unobservable. 

Conversely, the distinguishing characteristic of skill optimization is that every LLM rollout yields an explicit, structured, and semantically rich trajectory containing historical trial logs, intermediate planning rationales, and runtime error messages. These trajectories encapsulate the exact causal chains of success or failure. From this perspective, skill optimization shifts from black-box function querying into a stochastic coding problem, where the skill library serves as a repository of high-level software programs written in natural language. The LLM acts not merely as an evaluator, but as the underlying compiler and execution runtime, while the generated rollout trajectories function as the intermediate program execution traces. 

Traditional ZO optimization is forced to approximate gradients blindly through numerical variations because it cannot inspect the underlying function. In contrast, skill optimization leverages readable execution traces to perform targeted, semantic-driven debugging. In Section \ref{sec:skillopt-lite}, we introduce SkillOpt-Lite, a framework explicitly engineered to exploit this program-compiler paradigm by converting semantic execution traces into actionable code patches.

\begin{insightbox}
    \textbf{Insight 1 (Conceptual Divergence):} Classical ZO optimization relies on blind numerical perturbations, whereas agentic skill optimization functions as language-mediated program compilation where rollout trajectories serve as interpretable debugging feedback.
\end{insightbox}

\subsection{Generalization Error: A PAC-Learning Perspective}

Despite its formulation as a language-mediated coding process, skill optimization remains a statistical learning problem governed by the foundational principles of PAC-learning. Under the PAC framework, the generalization error $\epsilon(\mathcal{S})$ of a learned skill library $\mathcal{S}$ over the true task distribution $\mathcal{D}$ is bounded by its empirical error $\hat{\epsilon}_D(\mathcal{S})$ plus a compliance penalty. Instead of relying on uniform capacity measures of the static hypothesis space, this generalization gap can be tightly bounded via the lens of Expected On-Average Stability~\cite{shalev2010learnability}:
$$\epsilon(\mathcal{S}) \le \hat{\epsilon}_D(\mathcal{S}) + \mathcal{O}\left(\beta_{\exp} + \sqrt{\frac{\ln(1/\delta)}{N}}\right)$$
where $N$ is the number of training instances, and $\beta_{\exp}$ denotes the expected stability coefficient of the skill refinement algorithm. 

Let $D = \{z_1, \dots, z_N\}$ be the training dataset, and let $D^{\setminus i}$ be the adjacent dataset formed by removing the $i$-th task instance $z_i$. An optimization algorithm $\mathcal{A}$ exhibits Expected On-Average Stability with a coefficient $\beta_{\exp}$ if the expectation of its performance variance over the true data distribution $\mathcal{D}$ satisfies:
$$\mathbb{E}_{D, z \sim \mathcal{D}} \left| R(H(M, z, \mathcal{A}(D))) - R(H(M, z, \mathcal{A}(D^{\setminus i}))) \right| \le \beta_{\exp}$$
In the semantic domain, $\beta_{\exp}$ measures the algorithm's statistical resistance to episodic idiosyncrasies. If the refinement process overfits to a specific sample anomaly—such as hardcoding case-by-case branching actions or mimicking local environment variables unique to a single failed trial—the stability coefficient $\beta_{\exp}$ increases, leading to generalization collapse. To minimize $\beta_{\exp}$, the optimization algorithm must discard single-sample eccentricities and extract the common attributes across heterogeneous rollouts. This stabilizing compression forces the evolving skills to capture invariant structural logic rather than memorizing single-trial trajectories. 

Practical frameworks enforce cross-task consensus through various designs: Trace2Skill~\cite{ni2026trace2skill} and SkillForge~\cite{liu2026skillforge} aggregate rollouts into mini-batches or batch ticket pools to extract invariant lessons, whereas SkillOpt~\cite{yang2026skillopt} executes a hierarchical tree reduction via parallelized LLM merging.

\begin{insightbox}
    \textbf{Insight 2 (Stability and Overfitting):} From the lens of algorithmic stability, a robust skill optimization algorithm acts as a $\beta_{\exp}$-stabilizing operator, ensuring that text-based skill mutations remain invariant to single-trial rollout anomalies to capture cross-task structural invariants.
\end{insightbox}

To select the optimal model during self-evolution, validation-based model selection is employed. Assuming an evaluation metric bounded by $M$, the generalization error of the empirically selected skill library complies with a standard model selection bound over $m$ validation samples~\cite{zhang_2023_ltbook}:
$$\epsilon(\mathcal{S}_{\text{val}}) \le \hat{\epsilon}_{\text{val}}(\mathcal{S}_{\text{val}}) + \mathcal{O}\left(\sqrt{\frac{\ln(1/\delta)}{m}}\right)$$
By contrasting these two generalization boundaries, we uncover a clear statistical dividend unlocked by the validation protocol. In the standard training bound, the generalization gap is anchored to the algorithm's single-sample sensitivity $\beta_{\exp}$. As the agent searches through an expressively dense natural language space, the tendency to fit individual sample features penalizes the generalization capacity. Crucially, once a disjoint validation set is introduced, the stability coefficient $\beta_{\exp}$ is completely removed from the model selection upper bound. However, the statistical validity of this decoupling hinges upon a fundamental rule: the validation set must be strictly disjoint and independent from the training set to yield an unbiased estimate of the generalization gap.

An evaluation of recent literature reveals a pervasive violation of this validation protocol: works like Reflexion~\cite{shinn2023reflexion} bypass dynamic validation entirely in an open loop, while frameworks like SkillCat~\cite{chen2026skillcat}, SkillAdapter~\cite{yu2026skilladaptor}, and Trace2Skill~\cite{ni2026trace2skill} compromise the validation bound by executing their gates either on direct clones of the source training failure instances or on sub-sampled training subsets.

\begin{insightbox}
    \textbf{Insight 3 (Independent Validation Set):} To guarantee generalization, a validation set must satisfy a dual mandate: strict statistical independence from the training data to prevent evaluation bias, and a sufficient sample size $m$ to suppress the variance bound $\mathcal{O}(\sqrt{\ln(1/\delta)/m})$.
\end{insightbox}
\section{SkillOpt-Lite}

\subsection{A Motivating Example: Writing Skills with Coding Agents}

As discussed in Section \ref{sec:diff}, skill optimization can be modeled as a program-compiler paradigm, where the skill document serves as the software program and the LLM acts as the underlying compiler. To evaluate this paradigm, we conducted a pilot study mimicking workflows where AI engineers leverage commercial coding assistants to refine agentic skills. Specifically, we collected the raw rollout trajectories generated by the initial batch of a GPT-5.4-nano optimization loop and stored them as local text files. We then deployed GitHub Copilot~\cite{githubcopilot} to autonomously explore this directory solely via primitive file system tools, with the explicit directive to diagnose systematic failure patterns across trajectories and apply minimal modification patches directly to the baseline skill files. 

Crucially, the assistant bypassed all pre-defined mathematical topologies—such as the mini-batch slicing or hierarchical tree-reduction operators utilized in SkillOpt—and committed these edits directly without executing any intermediate validation loops. The refined skills were subsequently evaluated on the downstream test sets of Spreadsheet~\cite{ma2024spreadsheetbench}, ALFWorld~\cite{shridhar2020alfworld}, LiveMath~\cite{he2026livemathematicianbench}, and DocVQA~\cite{mathew2021docvqa}.

\begin{figure}[htbp]
    \centering
    \includegraphics[width=0.55\textwidth]{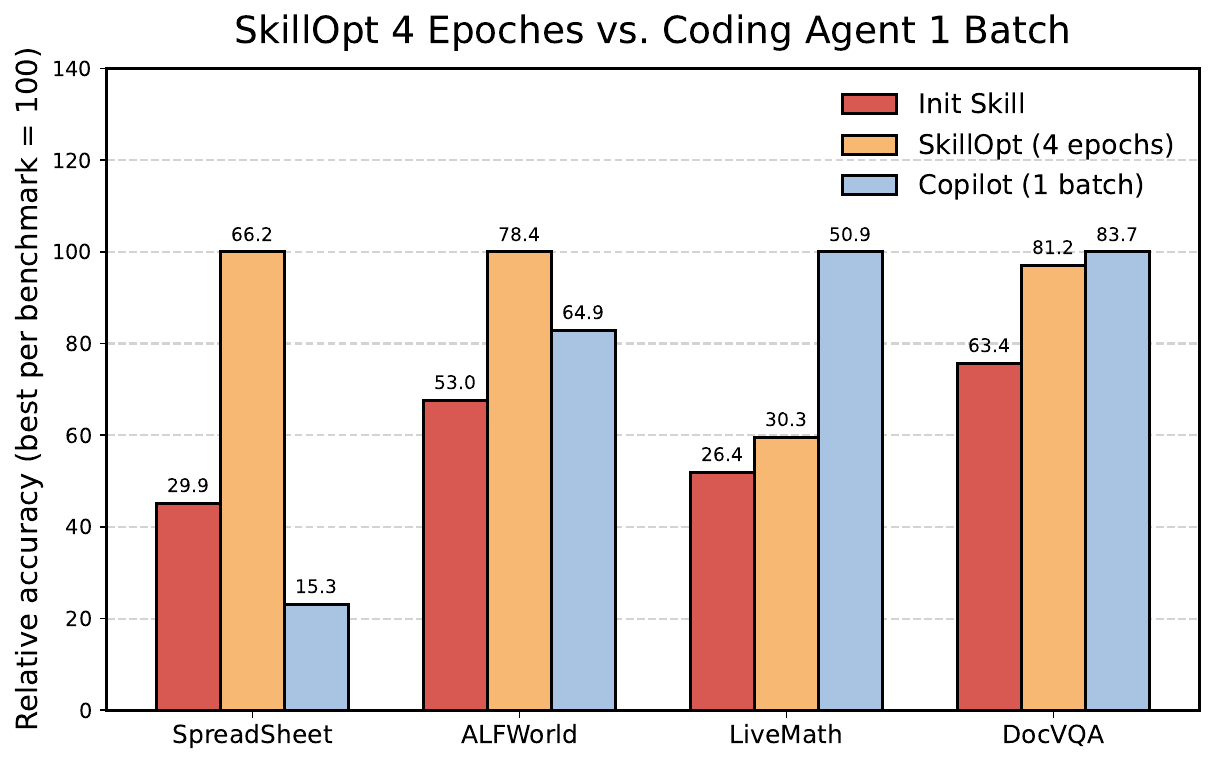}
    \caption{Empirical evaluation of single-batch coding agent exploration versus multi-epoch skill optimization.}
    \label{fig:skilloptvscopilot}
\end{figure}

The empirical results, illustrated in Figure \ref{fig:skilloptvscopilot}, reveal an instructive phenomenon. Without iterative loops, the coding assistant's single-batch file-system exploration achieved substantial performance gains. Notably, on both LiveMath and DocVQA, this single-batch unvalidated agent configuration outperformed the baseline SkillOpt framework even after the latter underwent four full epochs of iterative optimization. This trend highlights the capability of language models when operating directly over raw execution files. Conversely, on the Spreadsheet benchmark, the optimized skill degraded below its initial baseline performance, demonstrating the necessity of the closed-loop validation mechanisms embedded within the formal optimization pipeline.

\begin{insightbox}
    \textbf{Insight 4 (A Bitter Lesson in Skill Optimization):} As base models scale, complex algorithmic pipelines designed for skill updates can be consistently matched or outperformed by treating system artifacts as flat files and granting the model primitive file-system navigation tools.
\end{insightbox}

\subsection{SkillOpt-Lite Pipeline}
\label{sec:skillopt-lite}
\begin{figure}[htbp]
    \centering
    \begin{subfigure}{\textwidth}
        \centering
        \includegraphics[width=\textwidth]{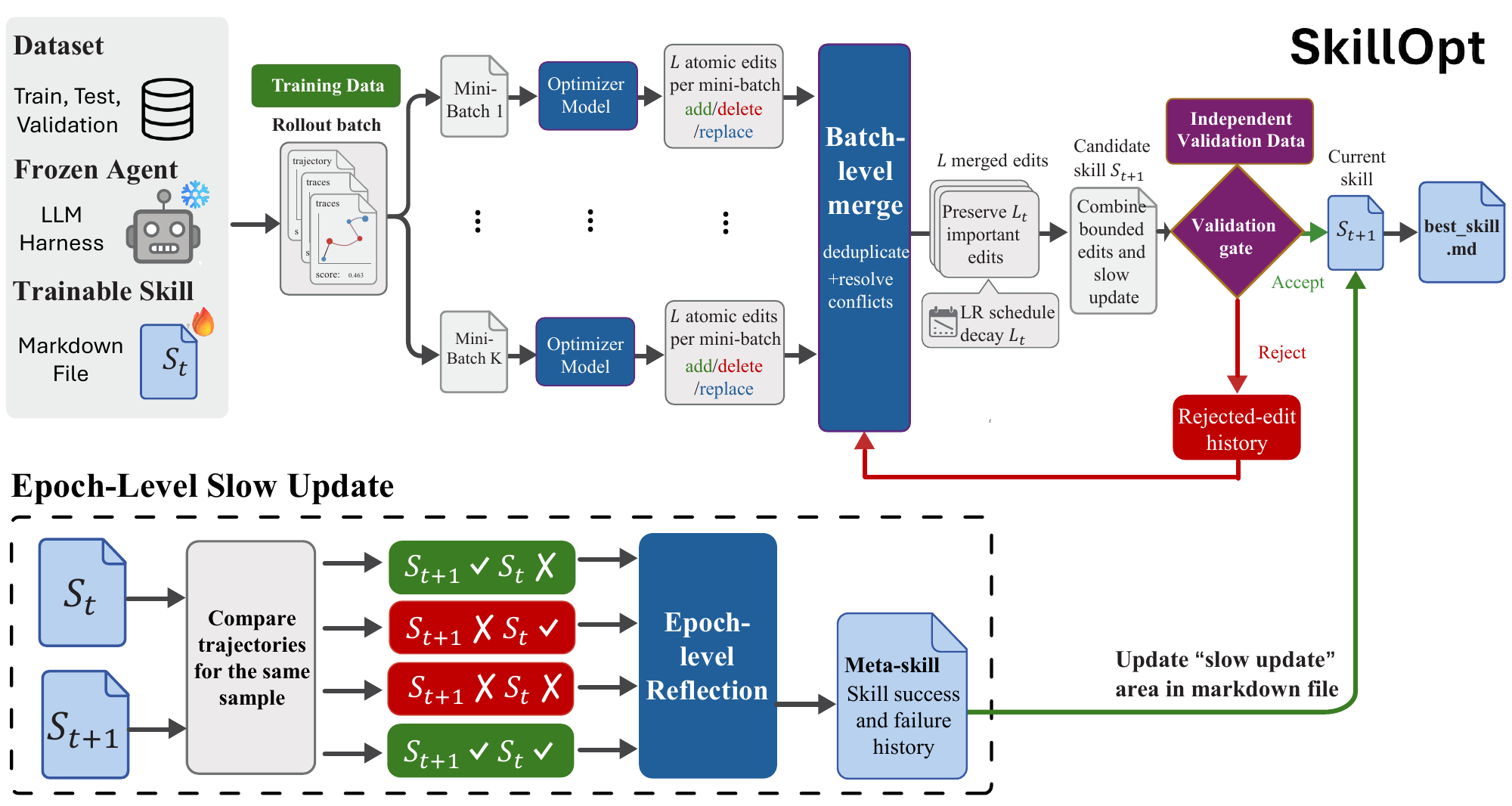}
        \caption{The original SkillOpt pipeline \cite{yang2026skillopt}.}
        \label{fig:pipeline_original}
    \end{subfigure}
    
    \vspace{10pt} 
    
    \begin{subfigure}{\textwidth}
        \centering
        \includegraphics[width=\textwidth]{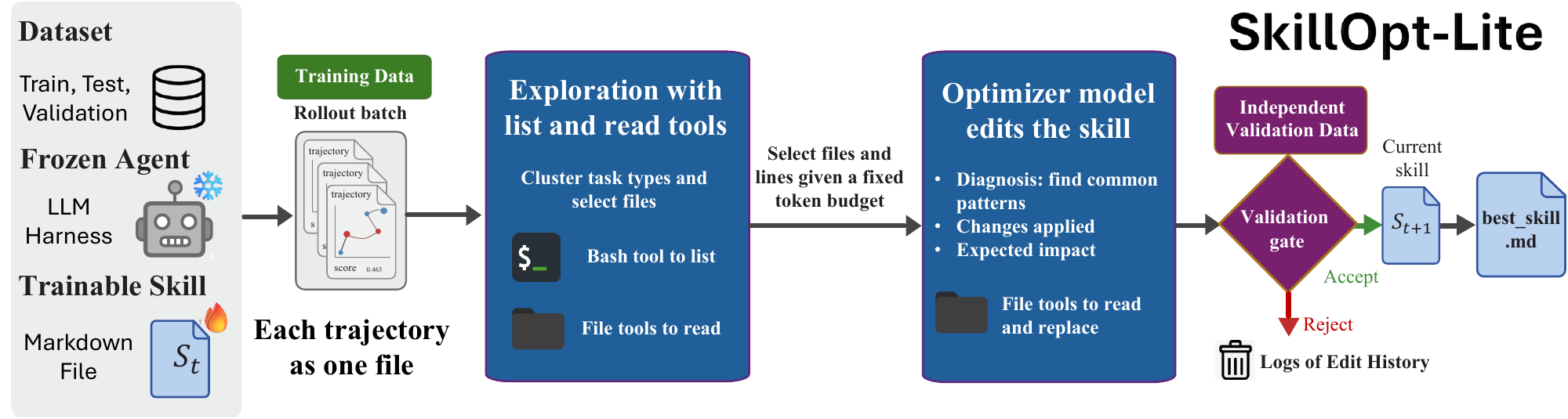}
        \caption{The proposed SkillOpt-Lite pipeline. The framework treats rollout trajectories as independent text files, allowing the optimizer model to explore the directory and extract shared failure patterns. Compared to the baseline SkillOpt framework, our streamlined design eliminates mini-batch reflection pooling, slow update damping, and historical rejection buffers.}
        \label{fig:pipeline_lite_sub}
    \end{subfigure}

\vspace{10pt}
    \begin{subfigure}{\textwidth}
        \centering
        \includegraphics[width=\textwidth]{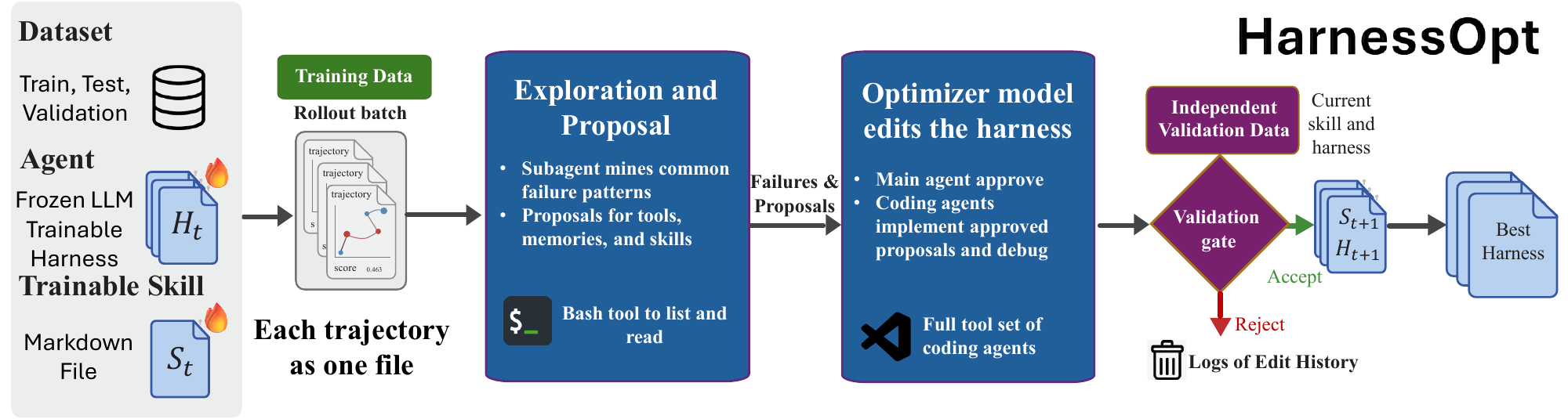}
        \caption{The proposed HarnessOpt pipeline. Because SkillOpt-Lite abstracts all artifacts as flat files, the operational harness naturally reduces to editable code files. Consequently, by deploying a coding agent as the core optimizer, the minimal skill pipeline natively generalizes into a full-scale harness optimization framework.}
        \label{fig:pipeline_harness_sub}
    \end{subfigure}
    
    \vspace{6pt} 
    \caption{Structural comparison of the optimization pipelines.}
    \label{fig:skillopt_lite}
\end{figure}

We propose \textbf{SkillOpt-Lite}, a streamlined framework designed to realize a minimal viable pipeline for skill optimization. As illustrated in the structural comparison in Figure \ref{fig:skillopt_lite}, we remove complex batch-level merging, text-based learning rate scheduling, and historical rejected-edit buffers. Furthermore, we deprecate the epoch-level slow update mechanism, which previously required matching and comparing trajectories across multiple epochs for meta-skill reflection. 

Instead, SkillOpt-Lite operates directly on the local disk via an autonomous debugging loop. The rollout batch is decoupled such that each individual trajectory is stored as a standalone file, enabling the system to navigate, read, and refine the skill library using native file-system tools. The workflow of SkillOpt-Lite is structured as follows:

\begin{description}
    \item[\textbf{1. Trajectory Staging:}] Following each batch rollout of the frozen agent harness, the raw trajectory—encompassing planning rationales, environment states, and task scores—is dumped directly onto the disk as a text file. As shown in Figure~\ref{fig:skillopt_lite}\subref{fig:pipeline_lite_sub}, each trajectory is stored as an independent log file rather than being aggregated into structural mini-batches.
    
    \item[\textbf{2. Trajectory Exploration:}] Rather than loading the entire corpus of raw logs into the model context window, the system utilizes explicit file-system utilities to navigate the workspace, adhering to the design principle outlined in Insight 4. Operating under a constrained token budget, the optimizer model executes autonomous shell commands to list directories, cluster uniform task failures, and select high-leverage files for detailed inspection.
    
    \item[\textbf{3. Consensus Mining and Minimal Edit:}] Equipped with file-reading capabilities, the optimizer reads these trajectory files directly from the directory to discover cross-task invariants and shared failure patterns, fulfilling the objective of Insight 2. Once these patterns are identified, and to satisfy the trust-region constraint of bounded text updates, the system follows a Minimal Update Principle by generating a compact code diff or patch to address the diagnosed errors.
    
    \item[\textbf{4. Validation Gating:}] The proposed patch is applied to construct a candidate skill library $\tilde{\mathcal{S}}$, which is immediately evaluated on an independent validation set to secure an unbiased empirical score, implementing the verification mechanism of Insight 3. If the candidate skill improves upon the current baseline score, it is accepted as the active skill $\mathcal{S}_{t+1}$. If it additionally exceeds the historical best score, it permanently overwrites the production skill file on disk, designated as \texttt{best\_skill.md}. Rejected updates are shunted directly to historical logs for archiving.
\end{description}

To demonstrate practical utility, we encapsulate this pipeline into a production-ready VS Code extension. Developers can trigger the full optimization loop directly within their development environment via a single-line slash command:

\begin{tcolorbox}[
    colback=vscodedark, 
    colframe=vscodeframe, 
    coltext=vscodetext,
    arc=4pt, 
    boxrule=0.8pt, 
    top=6pt, 
    bottom=6pt, 
    left=8pt, 
    right=8pt
]
    \small\ttfamily /skillopt-loop rounds=10 batchsize=40 target=gpt5.4-nano [custom\_requirements]
\end{tcolorbox}

This integration eliminates the configuration overhead typical of multi-agent platforms, delivering an interactive, lightweight debugging utility directly embedded inside the native IDE environment.
\section{Experiments}
\label{sec:experiments}

\subsection{Performance}
Our experimental configuration follows the evaluation protocols established in SkillOpt, utilizing SearchQA~\cite{dunn2017searchqa}, Spreadsheet~\cite{ma2024spreadsheetbench}, ALFWorld~\cite{shridhar2020alfworld}, LiveMath~\cite{he2026livemathematicianbench}, DocVQA~\cite{mathew2021docvqa}, and OfficeQA~\cite{opsahl2026officeqa}. We employ identical optimizer models for both SkillOpt and SkillOpt-Lite, with GitHub Copilot deployed as the baseline coding agent. Under the original dataset splits, LiveMath and OfficeQA contain only 10 to 20 validation instances, resulting in high validation variance. To mitigate this empirical instability, we adjust the train-validation-test split for these two benchmarks from $2:1:7$ to $2:2:6$. For OfficeQA, we implement an offline evaluation variant due to the absence of active search tool APIs. 

\begin{table}[htbp]
\centering
\small
\caption{Main experimental results across varying model scales. Subscripts denote the relative performance alignment compared with the initial baseline (\texttt{Init skill}). Bold text highlights the best performance within each model block. Light blue rows highlight the performance of our proposed \texttt{SkillOpt-Lite}. SkillOpt is executed for $\max$(4 epochs, 10 batches), while SkillOpt-Lite is evaluated strictly over 10 batches.}
\label{tab:main_results_final_verified}
\resizebox{\textwidth}{!}{%
\begin{tabular}{llcccccc}
\toprule
\textbf{Model} & \textbf{Skill source} & \textbf{SearchQA} & \textbf{Spreadsheet} & \textbf{ALFWorld} & \textbf{LiveMath} & \textbf{\shortstack{OfficeQA \\ (Offline)}} & \textbf{DocVQA} \\
\midrule
\multirow{3}{*}{GPT-4o} 
& Init skill & 64.6 & 44.1 & 82.3 & 25.9 & 21.0 & 78.3 \\
& SkillOpt   & 70.1\gain{5.5} & 48.7\gain{4.6} & 95.3\gain{13.0} & 31.2\gain{5.3} & 30.2\gain{9.2} & 85.2\gain{6.9} \\
\rowcolor{liteblue}
& SkillOpt-Lite & \textbf{71.5}\gain{6.9} & \textbf{49.8}\gain{5.7} & \textbf{97.8}\gain{15.5} & \textbf{58.8}\gain{32.9} & \textbf{32.4}\gain{11.4} & \textbf{86.4}\gain{8.1} \\
\midrule
\multirow{3}{*}{GPT-5.4-nano} 
& Init skill & 50.9 & 29.9 & 34.3 & 26.4 & 10.8 & 63.4 \\
& SkillOpt   & \textbf{68.8}\gain{17.9} & 51.6\gain{21.7} & 71.8\gain{37.5} & 30.3\gain{3.9} & \textbf{18.4}\gain{7.6} & 80.4\gain{17.0} \\
\rowcolor{liteblue}
& SkillOpt-Lite & 66.9\gain{16.0} & \textbf{66.2}\gain{36.3} & \textbf{81.3}\gain{47.0} & \textbf{55.7}\gain{29.3} & 15.5\gain{4.7} & \textbf{82.1}\gain{18.7} \\
\midrule
\multirow{3}{*}{GPT-5.4-mini} 
& Init skill & 69.6 & 28.2 & 82.3 & 34.9 & 23.0 & 85.3 \\
& SkillOpt   & 72.3\gain{2.7} & 47.5\gain{19.3} & \textbf{100.0}\gain{17.7} & 41.2\gain{6.3} & \textbf{32.1}\gain{9.1} & 90.9\gain{5.6} \\
\rowcolor{liteblue}
& SkillOpt-Lite & \textbf{73.1}\gain{3.5} & \textbf{73.3}\gain{45.1} & \textbf{100.0}\gain{17.7} & \textbf{63.2}\gain{28.3} & 30.4\gain{7.4} & \textbf{92.5}\gain{7.2} \\
\midrule
\multirow{3}{*}{GPT-5.4} 
& Init skill & 68.3 & 39.9 & 88.1 & 46.2 & 29.7 & 87.7 \\
& SkillOpt   & \textbf{77.5}\gain{9.2} & 61.5\gain{21.6} & \textbf{100.0}\gain{11.9} & 54.0\gain{7.8} & 40.2\gain{10.5} & 90.3\gain{2.6} \\
\rowcolor{liteblue}
& SkillOpt-Lite & 76.3\gain{8.0} & \textbf{79.4}\gain{39.5} & \textbf{100.0}\gain{11.9} & \textbf{66.0}\gain{19.8} & \textbf{45.3}\gain{15.6} & \textbf{91.2}\gain{3.5} \\
\midrule
\multirow{3}{*}{GPT-5.5} 
& Init skill & 72.4 & 37.4 & 90.1 & 36.6 & 33.2 & 89.0 \\
& SkillOpt   & 78.8\gain{6.4} & 76.2\gain{38.8} & \textbf{100.0}\gain{9.9} & 64.8\gain{28.2} & 72.2\gain{39.0} & 91.2\gain{2.2} \\
\rowcolor{liteblue}
& SkillOpt-Lite & \textbf{79.0}\gain{6.6} & \textbf{79.7}\gain{42.3} & \textbf{100.0}\gain{9.9} & \textbf{73.6}\gain{37.0} & \textbf{76.2}\gain{43.0} & \textbf{94.2}\gain{5.2} \\
\bottomrule
\end{tabular}%
}
\end{table}

\noindent
\textbf{Significant Improvements on Reasoning Tasks (LiveMath \& Spreadsheet) --} On tasks requiring complex reasoning and deterministic code execution, SkillOpt-Lite demonstrates substantial performance gains. For instance, on LiveMath, absolute accuracy increases from 31.2 to 58.8 (+27.6 points) for GPT-4o, and from 36.6 to 73.6 (+37.0 points) for GPT-5.5 relative to the baseline. On Spreadsheet, the streamlined pipeline similarly outperforms the full SkillOpt framework by a clear margin (e.g., 79.4 vs. 61.5 points on GPT-5.4). 

Mechanistically, this divergence occurs because the baseline framework relies on mini-batch reflection pooling, an operation that averages multiple distinct textual updates and consequently blurs the implicit gradient signal within discrete language spaces. By eliminating this averaging artifact, SkillOpt-Lite empowers the optimizer model to issue localized, discrete file-system edits that directly resolve logical and algorithmic deadlocks within the skill codebase.

\vspace{0.5em}
\noindent
\textbf{Consistent Bounds on Semantics-Heavy Domains (SearchQA, ALFWorld \& OfficeQA) --} Conversely, for tasks dominated by open-ended text retrieval or physical environment grounding, the performance differentials between the two frameworks are marginal. Across SearchQA, ALFWorld, and OfficeQA, SkillOpt-Lite either matches or slightly exceeds the baseline framework, typically fluctuating within a $+0.1$ to $+1.5$ point window. This behavior stems from the nature of these benchmarks, which require broad semantic coverage rather than deep, multi-step algorithmic execution. 

Because the underlying models rapidly saturate the available optimization margin within these shallow text-action domains, both variants converge toward statistically similar local optima. Crucially, however, our streamlined version achieves this upper bound with significantly reduced computational overhead, demonstrating that complex optimization infrastructure remains largely redundant for shallow, semantic-heavy domains.

\subsection{Convergence Speed}
\label{sec:convergence}

\begin{figure}[htbp]
    \centering
    \includegraphics[width=\textwidth]{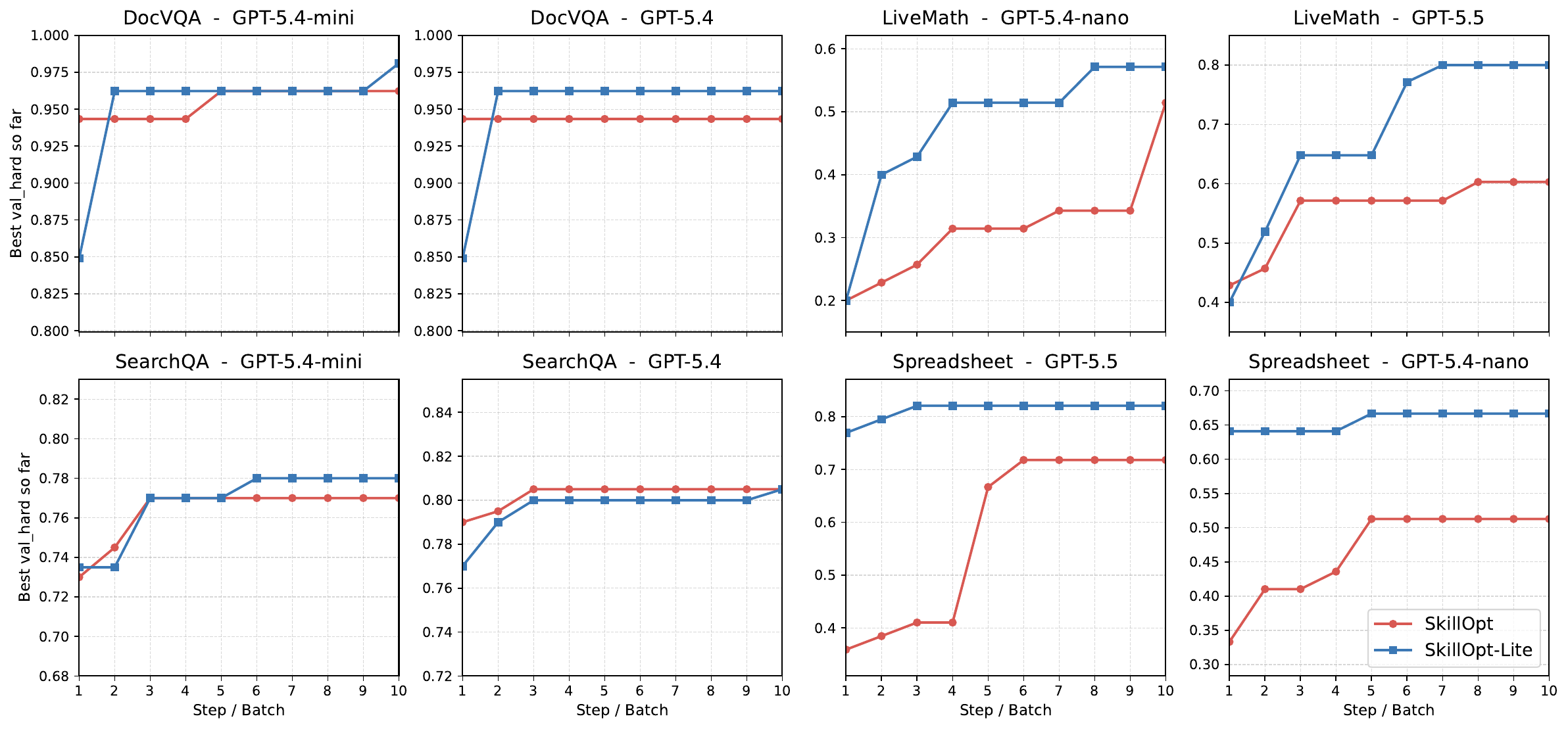}
    \caption{Convergence curve comparison between SkillOpt (red curve with circles) and SkillOpt-Lite (blue curve with squares) across various tasks and model scales over 10 optimization steps. The y-axis denotes the best validation performance achieved so far.}
    \label{fig:convergence_comparison}
\end{figure}

To evaluate the optimization efficiency of the proposed minimal pipeline, we analyze the convergence trajectories of SkillOpt and SkillOpt-Lite. Figure~\ref{fig:convergence_comparison} tracks the historical best validation scores ($\text{Best val}_{\text{hard}}\text{ so far}$) across 10 sequential optimization intervals under representative model scales and benchmarks.

The empirical trajectories highlight two distinct operational advantages of our streamlined framework:
\begin{itemize}
    \item \textbf{Accelerated Optimization Velocity:} Compared to the full SkillOpt framework, SkillOpt-Lite demonstrates a steeper initial optimization trajectory. For instance, in complex domains such as \texttt{LiveMath-GPT-5.5} and \texttt{LiveMath-GPT-5.4-nano}, SkillOpt-Lite secures substantial performance gains within the initial 2 to 3 steps. In contrast, the baseline SkillOpt framework exhibits sub-optimal trajectories in early phases, induced by the structural overhead of mini-batch partitioning and slow update damping mechanics.
    \item \textbf{Superior Performance:} Removing the multi-agent rejection buffer and reflection pooling does not induce premature convergence or destabilize policy evolution. Instead, SkillOpt-Lite consistently establishes an equal or elevated performance ceiling by the final optimization step. As observed in \texttt{Spreadsheet-GPT-5.4-nano} and \texttt{DocVQA-GPT-5.4}, the autonomous file-system exploration loop enables the optimizer model to isolate structural code defects directly, yielding precise refinements that bypass local sub-optimal states.
\end{itemize}

\section{Towards Harness Optimization}
\subsection{HarnessOpt Pipeline}

The primary motivation for extending our framework emerges when agentic skill optimization approaches an empirical performance saturation ceiling. Once declarative text prompts and heuristics are optimally tuned, further optimization gains must be unlocked by modifying the imperative control flows and tool execution scaffolding, known as harness optimization~\cite{ning2026code,lee2026meta,zhang2026self,harnessbenefit}. Because the core philosophical foundation of \textit{SkillOpt-Lite} treats all agent artifacts simply as standard file-system entities, this minimal pipeline can be natively extended to modify runtime codebase scripts. We designate this fully extended, unified framework as \textbf{HarnessOpt}. By lifting file-path restrictions within the workspace directory instead of a single skill markdown file, the optimizer model leverages the identical three-pillar architecture to directly refine the agent's execution routines. An illustration of the pipeline is shown in Fig. \ref{fig:pipeline_harness_sub}.

\noindent\textbf{Round-0 Bootstrapping and Human-in-the-Loop Gate --} 
In harness optimization, the structural integrity of the initial baseline framework---designated as the Round-0 harness---is critical. Blindly mutating raw control logic without a stable initial interface can trigger instant syntax collapse or broken tool dependencies. To address this risk, we implement a structured bootstrapping phase at Step 0. The framework first executes a full rollout of all training instances under the seed harness to aggregate raw system execution traces and exception logs. Next, a specialized diagnostic subagent scans these log directories to perform consensus mining. Its objective is to isolate cross-task architectural deficiencies rather than writing direct code edits, specifically identifying issues such as missing tool primitives or systemic API parsing mismatches. The subagent compiles these diagnostic insights into a structured configuration proposal. This proposal is evaluated through five critical lenses---tool inventory, prompt context, loop policy, codegen-shape, and state memory---culminating in three explicit decisions regarding memory integration, tool augmentation, and control flow shape. It is rendered as an interactive prompt within the VS Code environment, requiring explicit user approval before any structural alterations are committed to the core codebase.

\noindent\textbf{Automated Continuous Evolution and Sandboxed Gating --} 
Once the Round-0 harness is approved, \textbf{HarnessOpt} transitions into a fully automated, closed-loop continuous evolution phase where user intervention is bypassed. In rounds $t \ge 1$, the framework relaxes file permissions, allowing the underlying model to autonomously deploy standard shell tools to inspect execution states and apply targeted code patches to the harness scripts. Crucially, because modifying executable control flows carries the inherent risk of introducing infinite loops or regressions, strict validation and safety guardrails are enforced through three critical loop invariants:
\begin{itemize}
    \item \textbf{Allowlist Constraints}: Modifications are strictly confined to framework scaffolding scripts, while task-specific skills and internal configurations are treated as a read-only denylist to prevent drift.
    \item \textbf{Smoke and Validation Gating}: Validation gating is rigorously performed within an isolated subprocess sandbox. Candidate harnesses must pass a compile check and an early smoke test ($N=5$) before running the full validation set to measure actual performance gains.
    \item \textbf{Reversible Evolution with Toggles}: All code edits are fully reversible via \texttt{git reset} rollbacks, and significant modifications are wrapped behind environment variable feature toggles (initialized at module import time) to ensure clear bisection and future ablation handles.
\end{itemize}
The validation gate measures the final accuracy score against the historical best configuration. A statistical dead band $\Delta$ depending on the sample number is established to filter out stochastic variance. Patches yielding improvements below this deadband are accepted only if they demonstrate non-trivial progression on the continuous secondary soft metrics; otherwise, the state is rolled back to prevent codebase inflation with non-functional artifacts. 

To seamlessly integrate this capability into the developer workflow, we expose the harness optimization pipeline as a production-level slash command within our environment. Developers can initiate the continuous evolution loop by providing the targeted iteration rounds, the batch size for train rollouts, the target base model, and a reference skill file. A typical invocation follows the format below:

\begin{tcolorbox}[
    colback=vscodedark, 
    colframe=vscodeframe, 
    coltext=vscodetext,
    arc=4pt, 
    boxrule=0.8pt, 
    top=6pt, 
    bottom=6pt, 
    left=8pt, 
    right=8pt
]
    \small\ttfamily /harnessopt-loop rounds=2 batchsize=40 target=gpt5.4-nano skill=best\_skill\_nano.md [custom\_requirements]
\end{tcolorbox}

\subsection{Experimental Evaluation on Harness Optimization}

In our preliminary experiments, we observed that for the majority of the six datasets used in our skill optimization benchmark, the existing infrastructure was already sufficient, or harness optimization yielded minimal changes. Therefore, we focus our evaluation on \texttt{SpreadsheetBench} as a representative and challenging case study to analyze the efficacy of \textbf{HarnessOpt}. We conduct evaluations across four language models of varying capacities: \texttt{GPT5.4-nano}, \texttt{GPT5.4-mini}, \texttt{GPT5.4}, and \texttt{GPT5.5}. The detailed results are shown in Table~\ref{tab:harnessopt_results}, which corresponds to the empirical data in \text{image\_0d4db6.png}.

\begin{table}[htbp]
\centering
\caption{Optimization results on \texttt{SpreadsheetBench} across different LLM baselines.}
\label{tab:harnessopt_results}
\begin{tabular}{lcccc}
\hline
\textbf{Method} & \textbf{GPT5.4-nano} & \textbf{GPT5.4-mini} & \textbf{GPT5.4} & \textbf{GPT5.5} \\ \hline
Init skill & 0.2989 & 0.2821 & 0.3986 & 0.3737 \\
SkillOpt & 0.5160 & 0.4750 & 0.6150 & 0.7620 \\
SkillOpt-Lite & 0.6619 & 0.7331 & 0.7936 & 0.7972 \\ \hline
\textbf{HarnessOpt w.o. skill} & 0.7651 & 0.8114 & 0.8363 & 0.8363 \\
\textbf{HarnessOpt w. skill} & \textbf{0.7758} & \textbf{0.8256} & \textbf{0.8505} & \textbf{0.8577} \\ \hline
\end{tabular}
\end{table}

\noindent\textbf{Analysis of Model Behavioral Changes --}
By applying \textbf{HarnessOpt}, we identified distinct operational bottlenecks across different model tiers and fixed them with targeted harness modifications:
\begin{itemize}
    \item \textbf{\texttt{GPT5.4-mini} and \texttt{GPT5.4}}: For these two models, the baseline performance was largely limited by insufficient observation of the data files and weak final answer verification. \textbf{HarnessOpt} addressed this by expanding the visible range of the spreadsheet preview and introducing a dedicated step for the agent to introspect its final outputs. This successfully reduced parsing and formatting errors.
    \item \textbf{\texttt{GPT5.5} and \texttt{GPT5.4-nano}}: In contrast, these two models frequently encountered scenarios where their reasoning chains would get stuck in repetitive loops when a tool execution failed. To break this deadlock, \textbf{HarnessOpt} automatically intercepted these repetitive states and applied a fallback policy (\texttt{retry reasoning=low}), allowing the models to recover and complete the tasks.
\end{itemize}

\noindent\textbf{The Importance of Joint Optimization --}
As shown in Table~\ref{tab:harnessopt_results}, optimizing the harness codebase alone (\textit{HarnessOpt w.o. skill}) significantly boosts performance over purely prompt-based optimization (\textit{SkillOpt-Lite}), raising \texttt{GPT5.4-nano} from $0.6619$ to $0.7651$ and \texttt{GPT5.5} to $0.8363$. Notably, with harness optimization, the lightweight \texttt{GPT5.4-nano} achieves $0.7758$ (\textit{HarnessOpt w. skill}), outperforming the much larger \texttt{GPT5.5} model running under a basic harness with full prompt optimization (\textit{SkillOpt}, at $0.7620$). This capability inversion demonstrates that refining environmental scaffolding can empower a smaller model to surpass a frontier-class model in a suboptimal environment. Crucially, the best overall results are achieved only when we optimize both components simultaneously (\textit{HarnessOpt w. skill}), reaching $0.8505$ for \texttt{GPT5.4} and $0.8577$ for \texttt{GPT5.5}. This indicates that while skills provide strategic guidelines, the harness ensures a reliable execution environment; real-world performance relies on their co-design.

\section{Conclusion and Future Work}
\label{sec:conclusion}

\subsection{Conclusion}
In this report, we formalize agentic skill training through the lens of zeroth-order optimization and statistical learning theory. We show that complex multi-agent pooling and text update-damping mechanisms are largely redundant as base models scale. Guided by ``everything is file'' philosophy, we propose SkillOpt-Lite, a minimal viable pipeline that treats rollout trajectories as independent flat files and leverages autonomous coding agents for targeted, semantic-driven debugging. Empirically, SkillOpt-Lite accelerates optimization convergence and consistently matches or outperforms heavily engineered baselines across multiple benchmarks. Finally, we encapsulate this workflow into a native IDE extension, demonstrating that file-centric skill editing can naturally generalize to full harness optimization HarnessOpt.

\subsection{Future Work}
To move toward fully autonomous agent self-evolution, we outline four practical directions for future research:

\begin{itemize}[leftmargin=*]
    \item \textbf{Skill Optimization for Frontier Model Distillation:} While distilling proprietary frontier models into open-weight equivalents is a standard industrial practice, active anti-distillation defenses and sub-optimal generation boundaries often hinder data quality. Utilizing skill optimization to refine teacher agents across granular capabilities provides a structured framework for generating high-quality distillation corpora. However, because this paradigm couples dataset generation with the active optimization of downstream architectures, designing fast, compute-efficient evaluation protocols for data selection remains a key challenge.
    
    \item \textbf{Harness Template Database for Automated Optimization:} While this work explores basic programmatic modifications of the execution harness, scaling harness optimization requires a robust reference framework. Developing a curated library of minimal harness templates is essential to provide effective structural priors for coding agents. Furthermore, heterogeneous harnesses require adaptive sandboxing to ensure secure interaction. While current state-of-the-art agents (e.g., Claude Code~\cite{claudecode}) offer effective safety protocols for local software engineering environments, extending these containment mechanisms to internet-augmented or multimodally grounded execution domains remains unaddressed.
    
    \item \textbf{Harness Evolution as Continual Learning:} Given the high computational overhead of continuously fine-tuning foundational weights, operating directly on the non-parametric layer—specifically treating execution harnesses and domain skills as the evolving parameters—presents a promising alternative for lifelong learning \cite{weng2026learning}. A primary challenge under this framework lies in the structural alignment and fusion of independently evolved lineages of harnesses over extended temporal horizons. A direct extension of this research involves developing personalized agents, where the harness architecture continuously adapts to match long-term human preferences and idiomatic developer workflows.

    \item \textbf{Extending Optimization to Foundation Model Training:} While the base foundation models remain frozen in our current framework, this optimization pipeline can naturally extend to the model training phase itself. In our philosophy, since data collections are managed as standard files, a foundation model represents an intrinsic compilation of these files. By using our identical three-pillar architecture, the framework can establish a comprehensive benchmark to stress-test model boundaries, automatically edit web-scraping controllers or data-distillation prompts to gather targeted samples, and use this data to train an iteratively better model. This shifts the optimization boundary from scripts to model parameters, achieving full agent self-evolution. Interestingly, we have already pioneered semi-automated versions of this approach, such as dynamically scaling image-text pairs during the pre-training of our internal multimodal generation and understanding models, and curating specific video training data during our development of LLaVA-OneVision-2 \cite{an2026llava} (though these structural implementation details remained unhighlighted in the main paper). Fully automating this data-model-harness co-design loop without manual intervention remains a critical next step.
\end{itemize}

\bibliography{references}

@misc{anthropic2025claudecode,
  author       = {{Anthropic}},
  title        = {Claude Code: An Agentic Command Line Tool},
  year         = {2025},
  howpublished = {\url{https://docs.anthropic.com/en/docs/agents-and-tools/claude-code/overview}},
  note         = {Accessed: 2026-06-24}
}

@misc{openclaw2025,
  author       = {{OpenClaw Contributors}},
  title        = {OpenClaw: An Open-Source Agentic Command-Line Assistant},
  year         = {2025},
  howpublished = {\url{https://github.com/openclaw/openclaw}},
  note         = {GitHub Repository}
}

@misc{nous2026hermes,
  author       = {{Nous Research}},
  title        = {Hermes Agent},
  year         = {2026},
  howpublished = {\url{https://nousresearch.com/hermes3/}},
  note         = {Advanced Agentic and Function-Calling Foundation Models}
}

@article{yang2026skillopt,
  title={Skillopt: Executive strategy for self-evolving agent skills},
  author={Yang, Yifan and Gong, Ziyang and Huang, Weiquan and Yang, Qihao and Zhou, Ziwei and Huang, Zisu and Li, Yan and Gao, Xuemei and Dai, Qi and Liu, Bei and others},
  journal={arXiv preprint arXiv:2605.23904},
  year={2026}
}

@article{yang2024swe,
  title={Swe-agent: Agent-computer interfaces enable automated software engineering},
  author={Yang, John and Jimenez, Carlos and Wettig, Alexander and Lieret, Kilian and Yao, Shunyu and Narasimhan, Karthik and Press, Ofir},
  journal={Advances in Neural Information Processing Systems},
  volume={37},
  pages={50528--50652},
  year={2024}
}

@article{lee2026meta,
  title={Meta-harness: End-to-end optimization of model harnesses},
  author={Lee, Yoonho and Nair, Roshen and Zhang, Qizheng and Lee, Kangwook and Khattab, Omar and Finn, Chelsea},
  journal={arXiv preprint arXiv:2603.28052},
  year={2026}
}

@article{lin2026agentic,
  title={Agentic harness engineering: Observability-driven automatic evolution of coding-agent harnesses},
  author={Lin, Jiahang and Liu, Shichun and Pan, Chengjun and Lin, Lizhi and Dou, Shihan and Xi, Zhiheng and Huang, Xuanjing and Yan, Hang and Han, Zhenhua and Gui, Tao and others},
  journal={arXiv preprint arXiv:2604.25850},
  year={2026}
}

@article{ni2026trace2skill,
  title={Trace2skill: Distill trajectory-local lessons into transferable agent skills},
  author={Ni, Jingwei and Liu, Yihao and Liu, Xinpeng and Sun, Yutao and Zhou, Mengyu and Cheng, Pengyu and Wang, Dexin and Zhao, Erchao and Jiang, Xiaoxi and Jiang, Guanjun},
  journal={arXiv preprint arXiv:2603.25158},
  year={2026}
}

@article{shinn2023reflexion,
  title={Reflexion: Language agents with verbal reinforcement learning},
  author={Shinn, Noah and Cassano, Federico and Gopinath, Ashwin and Narasimhan, Karthik and Yao, Shunyu},
  journal={Advances in neural information processing systems},
  volume={36},
  pages={8634--8652},
  year={2023}
}

@article{wang2023voyager,
  title={Voyager: An open-ended embodied agent with large language models},
  author={Wang, Guanzhi and Xie, Yuqi and Jiang, Yunfan and Mandlekar, Ajay and Xiao, Chaowei and Zhu, Yuke and Fan, Linxi and Anandkumar, Anima},
  journal={arXiv preprint arXiv:2305.16291},
  year={2023}
}

@article{chen2026skillcat,
  title={SkillCAT: Contrastive Assessment and Topology-Aware Skill Self-Evolution for LLM Agents},
  author={Chen, Kunfeng and Zhong, Qihuang and Liu, Juhua and Du, Bo},
  journal={arXiv preprint arXiv:2606.13317},
  year={2026}
}

@article{tao2026softskill,
  title={SoftSkill: Behavioral Compression for Contextual Adaptation},
  author={Tao, Xijia and Teng, Yihua and Fu, Xinyu and Liu, Ziru and Chen, Kecheng and Zhao, Yuzhi and Zhang, Suiyun and Liu, Rui and Kong, Lingpeng},
  journal={arXiv preprint arXiv:2606.20333},
  year={2026}
}

@article{yu2026skilladaptor,
  title={SkillAdaptor: Self-Adapting Skills for LLM Agents from Trajectories},
  author={Yu, Zhuoyun and Xie, Xin and Yao, Wuguannan and Wang, Chenxi and Liang, Lei and Qi, Xiang and Deng, Shumin},
  journal={arXiv preprint arXiv:2606.01311},
  year={2026}
}

@article{liu2026skillforge,
  title={Skillforge: Forging domain-specific, self-evolving agent skills in cloud technical support},
  author={Liu, Xingyan and Luo, Xiyue and Li, Linyu and Huang, Ganghong and Liu, Jianfeng and Qiao, Honglin},
  journal={arXiv preprint arXiv:2604.08618},
  year={2026}
}

@article{harnessbenefit,
  title={Harness Updating Is Not Harness Benefit: Disentangling Evolution Capabilities in Self-Evolving LLM Agents},
  author={Self-Evolution, Harness},
  year={2026}
}

@book{zhang_2023_ltbook,
   title={Mathematical Analysis of Machine Learning Algorithms},
   author={Zhang, Tong},
   doi={10.1017/9781009093057},
   publisher={Cambridge University Press},
   place={Cambridge},
   year={2023}
}

@article{zhang2026self,
  title={Self-Harness: Harnesses That Improve Themselves},
  author={Zhang, Hangfan and Zhang, Shao and Li, Kangcong and Zhang, Chen and Chen, Yang and Zhang, Yiqun and Bai, Lei and Hu, Shuyue},
  journal={arXiv preprint arXiv:2606.09498},
  year={2026}
}

@article{liu2020primer,
  title={A primer on zeroth-order optimization in signal processing and machine learning: Principals, recent advances, and applications},
  author={Liu, Sijia and Chen, Pin-Yu and Kailkhura, Bhavya and Zhang, Gaoyuan and Hero III, Alfred O and Varshney, Pramod K},
  journal={IEEE Signal Processing Magazine},
  volume={37},
  number={5},
  pages={43--54},
  year={2020},
  publisher={IEEE}
}

@article{shalev2010learnability,
  title={Learnability and stability in the Vapnik-Chervonenkis sense},
  author={Shalev-Shwartz, Shai and Shamir, Ohad and Srebro, Nathan and Sridharan, Karthik},
  journal={The Annals of Statistics},
  volume={38},
  number={5},
  pages={2642--2696},
  year={2010},
  publisher={Institute of Mathematical Statistics}
}

@misc{githubcopilot,
  title={{GitHub Copilot}},
  author={{GitHub}},
  year={2021},
  howpublished={\url{https://github.com/features/copilot}},
  note={Accessed: 2026}
}

@misc{claudecode,
  title={{Claude Code Source Code}},
  author={{Anthropic}},
  year={2026},
  howpublished={\url{https://github.com/codeaashu/claude-code}},
}

@article{dunn2017searchqa,
  title={Searchqa: A new q\&a dataset augmented with context from a search engine},
  author={Dunn, Matthew and Sagun, Levent and Higgins, Mike and Guney, V Ugur and Cirik, Volkan and Cho, Kyunghyun},
  journal={arXiv preprint arXiv:1704.05179},
  year={2017}
}

@article{ma2024spreadsheetbench,
  title={Spreadsheetbench: Towards challenging real world spreadsheet manipulation},
  author={Ma, Zeyao and Zhang, Bohan and Zhang, Jing and Yu, Jifan and Zhang, Xiaokang and Zhang, Xiaohan and Luo, Sijia and Wang, Xi and Tang, Jie},
  journal={Advances in Neural Information Processing Systems},
  volume={37},
  pages={94871--94908},
  year={2024}
}

@article{shridhar2020alfworld,
  title={Alfworld: Aligning text and embodied environments for interactive learning},
  author={Shridhar, Mohit and Yuan, Xingdi and C{\^o}t{\'e}, Marc-Alexandre and Bisk, Yonatan and Trischler, Adam and Hausknecht, Matthew},
  journal={arXiv preprint arXiv:2010.03768},
  year={2020}
}

@article{he2026livemathematicianbench,
  title={LiveMathematicianBench: A Live Benchmark for Mathematician-Level Reasoning with Proof Sketches},
  author={He, Linyang and Yu, Qiyao and Dong, Hanze and Liao, Baohao and Xu, Xinxing and Goldblum, Micah and Bian, Jiang and Mesgarani, Nima},
  journal={arXiv preprint arXiv:2604.01754},
  year={2026}
}

@inproceedings{mathew2021docvqa,
  title={Docvqa: A dataset for vqa on document images},
  author={Mathew, Minesh and Karatzas, Dimosthenis and Jawahar, CV},
  booktitle={Proceedings of the IEEE/CVF winter conference on applications of computer vision},
  pages={2200--2209},
  year={2021}
}

@article{opsahl2026officeqa,
  title={Officeqa pro: An enterprise benchmark for end-to-end grounded reasoning},
  author={Opsahl-Ong, Krista and Singhvi, Arnav and Collins, Jasmine and Zhou, Ivan and Wang, Cindy and Baheti, Ashutosh and Oertell, Owen and Portes, Jacob and Havens, Sam and Elsen, Erich and others},
  journal={arXiv preprint arXiv:2603.08655},
  year={2026}
}

@article{ning2026code,
  title={Code as agent harness},
  author={Ning, Xuying and Tieu, Katherine and Fu, Dongqi and Wei, Tianxin and Li, Zihao and Bei, Yuanchen and Zou, Jiaru and Ai, Mengting and Liu, Zhining and Li, Ting-Wei and others},
  journal={arXiv preprint arXiv:2605.18747},
  year={2026}
}

@misc{weng2026learning,
  author       = {Jiayi Weng},
  title        = {Learning Beyond Gradients},
  month        = {May},
  year         = {2026},
  howpublished = {\url{https://trinkle23897.github.io/learning-beyond-gradients/}},
  note         = {Accessed: 2026-06-30}
}

@article{an2026llava,
  title={LLaVA-OneVision-2: Towards Next-Generation Perceptual Intelligence},
  author={An, Xiang and Xie, Yin and Tang, Feilong and Yan, Yunyao and Tan, Huajie and Zhu, Didi and Chen, Changrui and Zhao, Xiuwei and Qin, Bin and Yang, Kaicheng and others},
  journal={arXiv preprint arXiv:2605.25979},
  year={2026}
}
\bibliographystyle{unsrtnat}

\end{document}